\documentclass[%
 reprint,
 amsmath,amssymb,
 aps,
]{revtex4-1}

\usepackage{graphicx}
\usepackage{dcolumn}
\usepackage{bm}
\usepackage[                   
            pdfstartview=FitH,
            colorlinks, 
            pdfborder=001,   
            linkcolor=blue,
            anchorcolor=blue,
            citecolor=blue,
            urlcolor=blue
            ]{hyperref}
\usepackage[landscape,papersize={297.1mm,210mm},left=1.9cm,right=1.6cm,top=2.7cm,bottom=2.8cm]{geometry}
\begin{document}
\title{Measuring  Coherence with Entanglement Concurrence}

\author{Xianfei Qi}
\author{Ting Gao}
\email{gaoting@hebtu.edu.cn}
\affiliation {College of Mathematics and Information Science, Hebei
Normal University, Shijiazhuang 050024, China}
\author{Fengli Yan}
\email{flyan@hebtu.edu.cn}
\affiliation {College of Physics Science and Information Engineering, Hebei
Normal University, Shijiazhuang 050024, China}

\begin{abstract}
Quantum coherence is a fundamental manifestation of the quantum superposition principle. Recently, Baumgratz \emph{et al}. [\href{http://dx.doi.org/10.1103/PhysRevLett.113.140401}{ Phys. Rev. Lett. \textbf{113}, 140401 (2014)}] presented a rigorous framework to quantify coherence from the view of theory of physical resource. Here we propose a new valid quantum coherence measure which is  a convex roof measure, for a quantum system of arbitrary dimension, essentially using the generalized Gell-Mann matrices.  Rigorous proof shows that the proposed coherence measure, coherence concurrence, fulfills all the requirements dictated by the resource theory
of quantum coherence measures. Moreover, strong links between the resource frameworks of coherence concurrence and entanglement concurrence is derived, which shows that any degree of coherence with respect to some reference basis can be converted to entanglement
via incoherent operations.  Our work provides a clear quantitative and operational connection between coherence and entanglement based on two kinds of concurrence.  This new coherence measure, coherence concurrence, may also be beneficial to the study of quantum coherence.
\end{abstract}


\maketitle

\section{Introduction}
As a striking feature of the quantum mechanics, quantum coherence arising from the principle of quantum superposition is important in quantum physics. Quantum coherence is one of the fundamental features which mark the departure of quantum world from classical realm, and the origin for extensive quantum phenomena such as interference, lasers, superconductivity, and superfluidity. It is an essential ingredient for numerous physical phenomena such as quantum optics,  quantum thermodynamics \cite{PRL113.150402, SR6.22174} etc.
 The catalytic role of quantum superposition states when used in thermal operations was uncovered \cite{PRL113.150402}.
In  \cite{SR6.22174}, the authors showed that the physical realisation of optimal thermodynamic projection processes can come with a non-trivial thermodynamic work only for quantum states with coherences.

Quantum coherence is also regarded as a fundamental ingredient for quantum information processing tasks \cite{QCI.2010}.
However, the comprehensive formulation of the resource theory of coherence was only recently presented \cite{PRL113.140401}, where coherence was identified to be intuitive and easily computable measures of coherence by adopting the viewpoint of coherence as a physical resource.
 Following this seminal work, fruitful research has been done, some of which was mainly devoted to finding new appropriate measures of quantum coherence \cite{PRL113.170401,PRA91.042120,QIC15.1307,PRA92.022124,PRA93.012110}, or studying  maximally coherent states \cite{QIC15.1355,PRA93.032326}, the issue of ordering states with coherence measures \cite{QIP2016}, distribution of quantum coherence in multipartite systems \cite{PRL116.150504}, and the relation between coherence and other measures of quantumness \cite{PRL115.020403,PRA92.022112,SR5.10922,PRL116.160407,PRL117.020402}.
  Coherence has also been studied in the context of incoherent quantum operations \cite{arXiv1604v2,JPA50.045301,PRX7.011024}.

Quantum entanglement is  the main ingredient of the quantum speed-up in quantum computation and communication. The role of entanglement as a resource in quantum information has stimulated intensive research trying to unveil both its qualitative and quantitative aspects \cite{RMP80.517}.
In the theory of entanglement, concurrence is an important entanglement measure. Concurrence was first introduced in Ref.\cite{PRA54.3824} as an auxiliary tool to compute
the entanglement of formation  for Bell-diagonal two-qubit
states. Subsequently, Wootters and co-workers established concurrence as an entanglement measure for two-qubit states and derived computable formulas for concurrence and entanglement of formation in the two-qubit case \cite{PRL78.5022,PRL80.2245}.  Later, generalizations to bipartite higher-dimensional systems \cite{PRA64.042315} as well as for multipartite systems \cite{PRL93.230501} were proposed. Though many lower bounds for concurrence based on various approaches were obtained \cite{PRA67.052308,PRL95.040504,PRA74.050303,PRA75.052330,PRL109.200503,QIP2017}, exact formulas were derived only for two-qubit states \cite{PRL80.2245} and some highly symmetric states \cite{PRL85.2625,PRA64.062307,PRA67.012307}. Several meaningful efforts have also been spent in generalizing the notion of concurrence to obtain new forms of concurrences for detecting multipartite entanglement \cite{PRA83.062325,PRA86.062323,PRL112.180501}. For entanglement quantified by the concurrence, monogamy of multipartite quantum systems were well studied \cite{PRA61.052306,PRL96.220503,PRA78.012311}.

In quantitative coherence theory, considerable effort has been spent in developing many different coherence measures, while much less is known regarding the relations between these measures, and in particular, their connection to the resources they quantify. It is believed that---given a well-defined coherence measure---there should be a physical resource (defined through a protocol) that is quantified by this measure. Based on the distance measure, the $l_1$-norm of coherence and the relative entropy of coherence were quantified \cite{PRL113.140401}.     Intrinsic randomness measure (also called coherence of formation) was proposed essentially using the intrinsic randomness of measurement \cite{PRA92.022124}. It equals coherence cost, which is the minimal asymptotic rate of consuming maximally coherent pure state for preparing $\rho$ by incoherent operation \cite{PRL116.120404}.
From the viewpoint of physical resource, this coherence measure indicates the operational aspect of quantum coherence.

Both coherence and entanglement display quantumness of a physical system. Therefore, it is meaningful to study the interconversion between quantum coherence and entanglement.  In this paper we put forward a new valid quantum coherence measure via the generalized Gell-Mann matrices, and derive the amount of one resource emerges from the other.

This paper is organized as follows. In Sec.~\uppercase\expandafter{\romannumeral 2} we review the framework of coherence measures and introduce three valid coherence measures, i.e., relative entropy of coherence, the $l_{1}$-norm of coherence, and the intrinsic randomness measure. In Sec.~\uppercase\expandafter{\romannumeral 3}, we present a new coherence measure called coherence concurrence for  any dimensional quantum system based on the generalized Gell-Mann matrices, and prove that it is a good coherence measure. In Sec.~\uppercase\expandafter{\romannumeral 4}, we establish a relation for the interconversion between entanglement and coherence under incoherent operations based on coherence concurrence and entanglement concurrence.  Sec.~\uppercase\expandafter{\romannumeral 5} is outlook and conclusion.

\section{Review of coherence measures}

Before we state our main results, a review of the framework of coherence measures is necessary. Throughout the paper, we consider a general $d$-dimensional Hilbert space $\mathcal{H}$. Note that coherence is basis
dependent, we fix a particular basis, $\{|i\rangle\}_{i=1,\ldots,d}$, of the $d$-dimensional Hilbert space $\mathcal{H}$ in which we consider our quantum states. A state is called incoherent if it is diagonal in this fixed basis  and otherwise coherent. The set of all incoherent states is usually labelled as $\mathcal{I}\subset \mathcal{H}$. Hence, all density operators $\delta\in \mathcal{I}$ are of the form
\begin{equation}
\begin{aligned}
\delta=\sum_{i=1}^{d}\lambda_{i}|i\rangle\langle i|,
\end{aligned}
\end{equation}
where $\lambda_i$ are probabilities.

In the resource theory of coherence, free operations are given by the so-called incoherent operations. An incoherent operation  is defined by an incoherent completely positive trace preserving \text{\small(ICPTP)} map. An incoherent operation $\Lambda_{\text{\tiny ICPTP}}$  is a
 completely positive trace preserving map  such that
\begin{equation}
\begin{aligned}
\Lambda_{\text{\tiny ICPTP}}(\rho)=\sum_{n}K_{n}\rho K_{n}^{\dag},
\end{aligned}
\end{equation}
with the Kraus operators $K_{n}$ satisfying $\sum_{n}K_{n}^{\dag}K_{n}=I_d$ and $K_{n}\mathcal{I}K_{n}^{\dag}\subset \mathcal{I}$. For the case where measurement outcomes are retained, the state corresponding to outcome $n$ is given by $\rho_{n}=K_{n}\rho K_{n}^\dag/p_{n}$ and occurs with probability $p_{n}=\text{tr}[K_{n}\rho K_{n}^\dag]$.

A maximal coherent state (MCS) is one that can be used as a resource to  prepare any other state of the same dimension with certainty by means of incoherent operations only. The following state
\begin{equation}
\begin{aligned}
|\Psi_{d}\rangle=\frac{1}{\sqrt{d}}\sum_{i=1}^{d}|i\rangle
\end{aligned}
\end{equation}
is a MCS.
By applying the unitary incoherent
operations on $|\Psi_{d}\rangle$, a set of maximally coherent states is obtained \cite{PRA93.032326}
\begin{equation}
\begin{aligned}
S_{\text{MCS}}=\left\{\frac{1}{\sqrt{d}}\sum\limits_{j=1}^{d}\mathrm{e}^{\mathrm{i}\theta_{j}}|j\rangle\mid \theta_{1},\ldots,\theta_{d}\in [0,2\pi)\right\}.
\end{aligned}
\end{equation}

A rigorous framework for quantifying coherence was proposed in Ref.\cite{PRL113.140401}.  A coherence measure is a map $C$ from quantum states $\rho$ to  nonnegative  real numbers satisfying the following properties:

(C1) ~$C(\rho)\geq 0$ for all states $\rho$, and $C(\delta)=0$ if and only if $\delta$ is an incoherent state.

(C2) Monotonicity under incoherent operators.
(C2a) $C(\rho)$ is nonincreasing under incoherent operations, i.e.,  $C(\Lambda_{\text{\tiny ICPTP}}(\rho))\leqslant C(\rho)$ for arbitrary incoherent operations $\Lambda_{\text{\tiny ICPTP}}$ and states $\rho$.
(C2b) $C(\rho)$ is nonincreasing on average under selective incoherent operations, i.e.,
$\sum_{n}p_{n}C(\rho_{n})\leqslant C(\rho)$ for all incoherent operations $\Lambda_{\text{\tiny ICPTP}}$ and states $\rho$, where probabilities $p_{n}=\text{tr}[K_{n}\rho K_{n}^\dag]$, states $\rho_{n}=K_{n}\rho K_{n}^\dag/p_{n}$, and Krause operators $K_{n}$ obeying $\sum_{n}K_{n}^{\dag}K_{n}=I_d$ and $K_{n}\mathcal{I}K_{n}^{\dag}\subset \mathcal{I}$.

(C3) Nonincreasing under the mixing processes of the states (convexity), that is, $C(\rho)$ is a convex function of density matrices, i.e., $C(\sum_{n}p_{n}\rho_{n})\leqslant \sum_{n}p_{n}C(\rho_{n})$ for any set of states $\{\rho_n\}$ and any probability distribution $\{p_n\}$.

Conditions (C2b) and (C3) automatically imply condition (C2a) \cite{PRL113.140401}.

(C4) Only MCSs can achieve maximal value.

The additional requirement (C4) for coherence measure was proposed in \cite{PRA93.032326}.

We will introduce some valid coherence measures satisfying all the four requirements. In Ref.\cite{PRL113.140401},
two widely known coherence measures quantified by the minimum distance from $\rho$ to all the incoherent states based on two different distance measures were presented. One is the \emph{relative entropy of coherence}, based on the relative entropy,
\begin{equation}
\begin{aligned}
C_{\text{rel.ent}}(\rho)\equiv \mathop{\textrm{min}}\limits_{\delta\in \mathcal{I}}S(\rho\parallel \delta)=S(\rho_{\text{diag}})-S(\rho),
\end{aligned}
\end{equation}
where $S$ is the von Neumann entropy and $\rho_{\text{diag}}$ is the dephased state in reference basis $\{|i\rangle\}$, i.e, the state obtained from $\rho$ by deleting all off-diagonal entries. Another is \emph{$l_{1}$-norm of coherence}, based on the $l_{1}$ matrix norm,
\begin{equation}
\begin{aligned}
C_{l_{1}}(\rho)\equiv  \mathop{\textrm{min}}\limits_{\delta\in \mathcal{I}}\|\rho-\delta\|_{l_{1}}=\sum\limits_{i\neq j}|\langle i|\rho|j\rangle|,
\end{aligned}
\end{equation}
which is the sum of the absolute value of the off-diagonal entries of the quantum state.

A quantum coherence measure, the \emph{intrinsic randomness measure}, essentially using the intrinsic randomness, was proposed in \cite{PRA92.022124}. It is the first convex roof measure for coherence. For pure state,
\begin{equation}
\begin{aligned}
R_{I}(|\psi\rangle\langle \psi|)=S(\rho_{\text{diag}}),
\end{aligned}
\end{equation}
which equals the relative entropy of coherence $C_{\text{rel.ent}}(|\psi\rangle\langle \psi|)$. This coherence measure is extended to mixed state by the so-called convex roof construction
\begin{equation}
\begin{aligned}
R_{I}(\rho)=\mathop{\textrm{min}}\limits_{\{p_{i},|\psi_{i}\rangle\}} \sum\limits_{i} p_{i}R_{I}(|\psi_{i}\rangle),
\end{aligned}
\end{equation}
where $\rho=\sum_{i}p_{i}|\psi_{i}\rangle\langle\psi_{i}|$, $p_i\geqslant 0$ and $\sum_{i}p_{i}=1$.
The quantify in the equation above is also known as coherence of formation, and was studied in \cite{PRL116.120404}.

\section{Coherence concurrence}
In this section, a new quantum coherence measure named \textquotedblleft coherence concurrence\textquotedblright, which is  a convex roof measure, for a quantum system of arbitrary dimension, via the generalized Gell-Mann matrices, is presented.
 It fulfills not only the original four requirements (C1), (C2a), (C2b), and (C3)  of coherence measures but also the additional requirement (C4),   and thus it is a valid coherence measure.

The generalized Gell-Mann matrices (GGM) are the generators of $SU(d)$ defined as the following three different types of matrices \cite{PLA314.339,JPA41.235303}:

(i) $d(d-1)/2$ symmetric GGM
\begin{equation}
\begin{aligned}
\Lambda_{\text{s}}^{j,k}=|j\rangle\langle k|+|k\rangle\langle j|,~~~(1\leqslant j<k\leqslant d);
\end{aligned}
\end{equation}

(ii) $d(d-1)/2$ antisymmetric GGM
\begin{equation}
\begin{aligned}
\Lambda_{\text{a}}^{j,k}=-\text{i}|j\rangle\langle k|+\text{i}|k\rangle\langle j|,~~~(1\leqslant j<k\leqslant d);
\end{aligned}
\end{equation}

(iii) $(d-1)$ diagonal GGM
\begin{equation}
\begin{aligned}
\Lambda^{l}=\sqrt{\frac{2}{l(l+1)}}&\left(\sum\limits_{j=1}^{l}|j\rangle\langle j|-l|l+1\rangle\langle l+1|\right),\\
           &(1\leqslant l\leqslant d-1).
\end{aligned}
\end{equation}

We give a new expression of $C_{l_{1}}$ based on symmetric GGM.
First, we introduce a lemma.

\emph{Lemma.} Let $A$ be Hermitian. If $A$ is positive semidefinite, then all of its principal submatrices are positive semidefinite \cite{MA.2013}.

\emph{Proposition.} For a density matrix $\rho$, there is
\begin{equation}\label{Prop}
\begin{aligned}
C_{l_{1}}(\rho)&=2\sum\limits_{1\leq j<k\leq d}|\rho_{jk}|\\
               &=\sum\limits_{1\leq j<k\leq d}\left|\sqrt{\eta_{1}^{j,k}}-\sqrt{\eta_{2}^{j,k}}\right|,
\end{aligned}
\end{equation}
where $\eta_{1}^{j,k}$ and $\eta_{2}^{j,k}$ are the non-zero eigenvalues of the matrix $\rho\Lambda_{\text{s}}^{j,k}\rho^{*}\Lambda_{\text{s}}^{j,k}$, and $\rho^*$ denotes complex conjugation in the standard basis.

\emph{Proof.} We just need to prove that
\begin{equation}\label{Eq-1}
\begin{aligned}
2|\rho_{jk}|=\left|\sqrt{\eta_{1}^{j,k}}-\sqrt{\eta_{2}^{j,k}}\right|.
\end{aligned}
\end{equation}
After tedious but straightforward  computation, the eigenvalues of the matrix $\rho\Lambda_{\text{s}}^{j,k}\rho^{*}\Lambda_{\text{s}}^{j,k}$ are $(|\rho_{jk}|+\sqrt{\rho_{jj}\rho_{kk}})^{2}$, $(|\rho_{jk}|-\sqrt{\rho_{jj}\rho_{kk}})^{2}$, and zeros. According to Lemma,  the square roots of non-zero eigenvalues are $|\rho_{jk}|+\sqrt{\rho_{jj}\rho_{kk}}$, $\sqrt{\rho_{jj}\rho_{kk}}-|\rho_{jk}|$, which implies Eq.(\ref{Eq-1}), as required.

Next we present a new quantum coherence measure, coherence concurrence.

For a $d$-dimensional pure state $|\psi\rangle$, we define its coherence concurrence as
\begin{equation}\label{CoherenceConcurrenceDef}
\begin{aligned}
C(|\psi\rangle)=\sum\limits_{1\leq j<k\leq d}|\langle \psi|\Lambda_{\text{s}}^{j,k}|\psi^{*} \rangle|.
\end{aligned}
\end{equation}
 It is not difficult to derive that
 \begin{equation}\label{PureState C=C_li}
  C(|\psi\rangle)=\sum\limits_{1\leq j<k\leq d}|\langle \psi|\Lambda_{\text{s}}^{j,k}|\psi^{*} \rangle|=C_{\text{$l_{1}$}}(|\psi\rangle\langle\psi|).
 \end{equation}
 That is, the coherence concurrence equals $l_{1}$-norm of coherence for pure states. Then, coherence concurrence is extended to mixed state by convex roof construction
\begin{equation}
\begin{aligned}
C(\rho)=\mathop{\textrm{min}}\limits_{\{p_{i},|\psi_{i}\rangle\}} \sum\limits_{i} p_{i}C(|\psi_{i}\rangle),
\end{aligned}
\end{equation}
where the minimization is taken over all possible ensemble realizations $\rho=\sum_{i}p_{i}|\psi_{i}\rangle\langle\psi_{i}|$, $p_i\geqslant 0$ and $\sum_{i}p_{i}=1$. The decomposition attaining the minimum value is said to be the optimal decomposition.

\emph{Theorem 1.} Coherence concurrence is a valid coherence measure. That is, the coherence concurrence satisfies all the requirements (C1)-(C4) of coherence measures.

\emph{Proof.}  For pure state, coherence concurrence satisfies  (C1), (C2a), (C2b), and (C3), as  coherence concurrence equals $l_{1}$-norm of coherence, while $l_{1}$-norm of coherence fulfills (C1), (C2a), (C2b), and (C3) \cite{PRL113.140401}.  For mixed states, it is easy to see from the definition that the coherence concurrence satisfies the requirements (C1) and (C3). As for the requirement (C2b), it can be proven in a similar way as shown in Ref. \cite{PRA92.022124}. Coherence concurrence fulfills (C2a) since it satisfies (C2b) and (C3) \cite{PRL113.140401}. Motivated by the proof in Ref. \cite{PRA93.032326}, we prove that coherence concurrence also fulfills (C4). Detailed proof of theorem is shown in Appendix A.

\emph{Corollary 1} ~ (1) Coherence concurrence is not less than the $l_{1}$-norm of coherence for mixed state, i.e., $C(\rho)\geq C_{\text{$l_{1}$}}(\rho)$ for any state $\rho$.

(2) For the family $\rho$ of mixed states, a pure state mixed with white noise, there is $C(\rho)=C_{l_{1}}(\rho)$.

It follows directly from Theorem 1 and the definition of coherence concurrence.

The relation between coherence concurrence and other coherence measures is listed in Table \uppercase\expandafter{\romannumeral1.}

\begin{table*}
\caption{\label{tab:table}The relations among four coherence measures\footnote{$C$,  $C_{l_{1}}$, $R_{I}$, $C_{\text{rel.ent}}$ denote coherence concurrence, $l_{1}$-norm of coherence, intrinsic randomness measure, relative entropy of coherence, respectively.}.}
\begin{ruledtabular}
\begin{tabular}{ccccc}
&$C$&$C_{l_{1}}$&$R_{I}$&$C_{\text{rel.ent}}$\\ \hline
 Qubit pure state&$C$ &$C_{l_{1}}=C$ &$R_{I}=H(C)$\footnote{$H(C)$ labels $H\left(\frac{1+\sqrt{1-C^{2}}}{2}\right)$.}&$C_{\text{rel.ent}}=R_{I}=H(C)$ \\
 Qubit mixed state&$C$
 &$C_{l_{1}}= C$ &$R_{I}=H(C)$&$ $\\
 Qudit pure state&$C$&$C_{l_{1}}=C$
 &&$C_{\text{rel.ent}}=R_{I}$\\
 Qudit mixed state &$C$&$C_{l_{1}}\leqslant C$&$$&$ $\\
 \end{tabular}
\end{ruledtabular}
\end{table*}

\section{The relation between coherence and entanglement}
In this section, we establish the connection between two quantum resources, coherence and entanglement, via coherence concurrence and entanglement concurrence. First, we review the knowledge about entanglement concurrence $C_{E}$. For bipartite pure state $|\psi\rangle\in \mathcal{H}_{M}\otimes \mathcal{H}_{N}$, entanglement concurrence is defined by
\begin{equation}\label{ConcurrenceDef}
\begin{aligned}
C_{E}(|\psi\rangle)=\sqrt{2(1-\textrm{tr}\rho_{M}^{2})},
\end{aligned}
\end{equation}
where $\rho_{M}=\textrm{tr}_{N}(|\psi\rangle\langle\psi|)$. For mixed state $\rho$, the concurrence is given by the minimum average concurrence taken over all decompositions of $\rho$, the so-called convex roof construction,
\begin{equation}
\begin{aligned}
C_{E}(\rho)=\mathop{\textrm{min}}\limits_{\{p_{i},|\psi_{i}\rangle\}} \sum\limits p_{i}C(|\psi_{i}\rangle).
\end{aligned}
\end{equation}
The convex roof is notoriously hard to evaluate, but for two qubits mixed state, an exact formula was given \cite{PRL80.2245}
\begin{equation}\label{CE}
\begin{aligned}
C_{E}(\rho) = \textrm{max}\{\lambda_{1}-\lambda_{2}-\lambda_{3}-\lambda_{4},0\},
\end{aligned}
\end{equation}
with the numbers $\lambda_{i}~(i=1, 2, 3, 4)$ are the square roots of the eigenvalues of the non-Hermitian matrix $\rho(\sigma_y\otimes\sigma_y)\rho^*(\sigma_y\otimes\sigma_y)$ in nonincreasing order, where $*$ denotes complex conjugation in the standard basis and $\sigma_y$ is the Pauli matrix.

Next, we will discuss the relation between coherence and entanglement. The following theorems provide a strong link between entanglement concurrence $C_{E}$ and coherence concurrence $C$.

\emph{Theorem 2.} The amount of entanglement $C_{E}$ generated from a state $\rho^{S}$ via an incoherent operation $\Lambda^{SA}$, by attaching an ancilla system $A$ initialized in a reference incoherent state $|1\rangle\langle 1|^{A}$, is bounded above by its coherence concurrence $C$:
\begin{equation}\label{Th2}
\begin{aligned}
C_{E}(\Lambda^{SA}[\rho^{S}\otimes |1\rangle\langle 1|^{A}])\leqslant C(\rho^{S}).
\end{aligned}
\end{equation}

\emph{Proof.} The combination of
\begin{equation}\label{Th2-1}
C(\rho^{S})=C(\rho^{S}\otimes |1\rangle\langle 1|^{A})\geqslant C(\Lambda^{SA}[\rho^{S}\otimes |1\rangle\langle 1|^{A}]),
\end{equation}
and
\begin{equation}\label{Th2-2}
 C(\Lambda^{SA}[\rho^{S}\otimes |1\rangle\langle 1|^{A}])\geqslant C_{E}(\Lambda^{SA}[\rho^{S}\otimes |1\rangle\langle 1|^{A}]),
\end{equation}
implies  Ineq.(\ref{Th2}).
Detailed proof of  theorem is shown in Appendix B.

This implies that   the system-ancilla state $\Lambda^{SA}[\rho^{S}\otimes |1\rangle\langle 1|^{A}]$ for any incoherent operation $\Lambda^{SA}$ is separable if the initial state $\rho^S$ of a $d$-dimensional system $S$ is incoherent. Namely, entanglement can be generated by incoherent operations if the initial state   $\rho^S$ is coherent.

An even stronger link exists for qubit system. We prove that inequality (\ref{Th2}) can be saturated for the case that both the system and the ancilla system are  qubit systems.

\emph{Corollary 2.} For any qubit state $\rho^{S}$,  there exists
an incoherent operation $\Lambda^{SA}$ such that the entanglement concurrence of two-qubit state generated from $\rho^{S}$ via $\Lambda^{SA}$, by attaching an ancilla qubit system $A$ initialized in a reference incoherent state $|1\rangle\langle 1|^{A}$, equals the coherence concurrence of $\rho^{S}$.

\emph{Proof.} Assume that $\rho^{S}=\sum_{i,j=1}^{2}\rho_{ij}|i\rangle\langle j|$.  We choose two-qubit \text{\small CNOT} gate as needed incoherent operation $\Lambda^{SA}$. Note that  coherence concurrence $C(\rho^{S})=C_{l_1}(\rho^S)=2|\rho_{12}|$ for qubit state $\rho^{S}$  Ref.\cite{PRA92.022124} and   $C_{E}(\Lambda^{SA}[\rho^{S}\otimes |1\rangle\langle 1|^{A}])=2|\rho_{12}|$ by Eq.(\ref{CE}).  Thus, $C_{E}(\Lambda^{SA}[\rho^{S}\otimes |1\rangle\langle 1|^{A}])=C(\rho^{S})$ as required. The conclusion is proved.

This shows that the degree of coherence concurrence in the initial state of qubit system $S$ can be exactly converted to an equal degree of
entanglement concurrence between $S$ and the incoherent ancilla qubit $A$ by suitable incoherent operation,  CNOT gate.
That is,
 the mount of entanglement concurrence $C_{E}$ generated from a qubit state $\rho^{S}$ via an incoherent operation $\Lambda^{SA}$, by attaching an ancilla qubit system $A$ initialized in a reference incoherent state $|1\rangle\langle 1|^{A}$, reaches the  maximum value when incoherent operation $\Lambda^{SA}$ is two-qubit \text{\small CNOT} gate, which is also the coherence concurrence $C(\rho^{S})$.

Next we show that any degree of coherence with respect to some reference basis can be converted to entanglement via incoherent operations.

\emph{Theorem 3.} For an arbitrary  state $\rho^{S}$,  there exists
an incoherent operation $\Lambda^{SA}$ such that the entanglement concurrence of bipartite state generated from $\rho^{S}$ via $\Lambda^{SA}$, by attaching an ancilla system $A$ initialized in a reference incoherent state $|1\rangle\langle 1|^{A}$, has the following inequality relation with its coherence concurrence:
\begin{equation}\label{Th3}
\begin{aligned}
C_{E}(\Lambda^{SA}[\rho^{S}\otimes |1\rangle\langle 1|^{A}]) \geqslant\sqrt{\frac{2}{d(d-1)}}C(\rho^{S}).
\end{aligned}
\end{equation}
Here the dimension $d_A$ of the ancilla is not smaller than that of the system, $d_A\geq d$.

\emph{Proof.} First, we prove that this inequality is satisfied for pure state. Then, it is extended to the case of mixed state. Detailed proof of theorem can be found in Appendix C.

\emph{Corollary 3.} If $\rho^{S}$ is a maximal coherent state,  there exists
an incoherent operation $\Lambda^{SA}$ such that (\ref{Th3}) can be saturated.

The following result ( Theorem 2 in \cite{PRL115.020403} )  follows immediately from Theorem 2 and Theorem 3:

 A state $\rho^{S}$ can be converted to an entangled
state via incoherent operations if and only if $\rho^{S}$ is coherent.

The coherence of quantum states is basis dependent as well as the entanglement of states \cite{EPJD64.181, PRL87.077901}.
Quantum coherence is basis dependent by its definition, while entanglement is locally basis independent, i.e., entanglement is invariant under local unitary transformations.
States that are entangled with respect to a given
partition in subsystems can be separable with respect to
another partition \cite{PRL87.077901}.
However, entanglement usually change if a global unitary is applied. That is, via global unitary transformations we can switch from an entangled state to a separable state. For pure states, we can always switch unitarily between
separability and maximal entanglement. However,
for mixed states a minimal mixedness is required because the maximal mixed state $\frac{1}{d_1d_2}\sum_{i=1}^{d_1}\sum_{j=1}^{d_2} |ij\rangle\langle ij|$
 and a sufficiently small neighborhood is
separable for any factorization  \cite{EPJD64.181}, that is,  any unitary transformations can not change the separabilty of  the maximal mixed state.

Except the maximal mixed state$\frac{I_d}{d}$, being incoherent  in any basis, we can always switch between coherence and incoherence.
Since coherence is a basis dependent concept, a unitary operation in general changes the coherence of a given state.
Every state $\rho=\sum_{i=1}^d \lambda_i|\varphi_i\rangle\langle\varphi_i|$ can be unitarily transformed into an incoherent state $U\rho U^\dagger=\sum_{i=1}^d \lambda_i|i\rangle\langle i|$, where $U$ is a unitary operator such that $U|\varphi_i\rangle=|i\rangle$. Theorem 2 in \cite{quant-ph1612.07570} shows that any state being different from the maximal mixed state, can be unitarily transformed into a coherent state.

\section{Outlook and conclusion}

The new coherence measure, coherence concurrence, may raise many interesting problems. One can discuss whether $l_{1}$-norm of coherence and coherence concurrence coincide. It would be of great interest to study coherence distribution in multipartite quantum systems based on the coherence concurrence. An elegant equation connects coherence concurrence with intrinsic randomness measure for qubit system. More research is needed to further study the potential link between them for qudit system. The relation between coherence concurrence and other coherence measures is also needed to be further investigated.

In summary, a new coherence measure \textquotedblleft coherence concurrence\textquotedblright ~is presented for any dimensional quantum system based on the generalized Gell-Mann matrices. It satisfies all the requirements for a proper quantum coherence measure and is convex roof measure.  We show that any degree of coherence in the initial state of a quantum system $S$ can be converted to entanglement between $S$ and the incoherent ancilla $A$ by some incoherent operation. In addition, we establish the relation for the interconversion between coherence and entanglement based on coherence concurrence and entanglement concurrence.   As a counterpart of entanglement concurrence for coherence manipulation, we expect that coherence concurrence can have various applications in theory of quantum coherence similar to the concurrence in entanglement theory.

\begin{acknowledgments}
This work was supported by the National Natural Science Foundation
of China under Grant Nos: 11371005, 11475054, Hebei Natural Science Foundation
of China under Grant No:  A2016205145.
\end{acknowledgments}

\appendix

\section{Detailed proof of Theorem 1}

We will show that the coherence concurrence satisfies all the requirements (C1)-(C4) of a proper quantum coherence measures. Here, we only show how to prove (C2b) and (C4), the proofs for the other requirements are stated in the main text.

\subsection{Proof of (C2b)}

For the pure state, the monotonicity requirement of (C2b) is,
\begin{equation}\label{A1}
\begin{aligned}
C(|\psi\rangle)\geqslant \sum\limits_{n} p_{n}C(|\psi_{n}\rangle),
\end{aligned}
\end{equation}
where $|\psi_{n}\rangle=K_{n}|\psi\rangle/\sqrt{p_{n}}$, and $p_{n}=\text{tr}[K_{n}|\psi\rangle\langle\psi|K_{n}^{\dag}]$. It is obvious that this requirement is satisfied, because the coherence concurrence equals $l_{1}$-norm of coherence $C_{l_{1}}$ for pure state, and the monotonicity of which has been proved \cite{PRL113.140401}.

For a mixed state $\rho$, suppose that $\rho=\sum_{i}p_{i}|\psi_{i}\rangle\langle\psi_{i}|$ is the optimal decomposition that achieves the minimum value. That is,
\begin{equation}
\begin{aligned}
C(\rho)=\sum\limits_{i} p_{i}C(|\psi_{i}\rangle).
\end{aligned}
\end{equation}
It remains to prove that for incoherent operators $\Lambda_{\texttt{ICPTP}}$ there must be
\begin{equation}
C(\rho)\geqslant\sum\limits_{n} p_{n}C(\rho_{n}),
\end{equation}
where $\rho_{n}=K_{n}\rho K_{n}^{\dag}/p_{n}$ and $p_{n}=\text{tr}[K_{n}\rho K_{n}^{\dag}]$. Note that
\begin{equation}
\begin{aligned}
\rho_{n}&=\frac{K_{n}\rho K_{n}^{\dag}}{p_{n}}\\
        &=\sum\limits_{i}\frac{p_{i}}{p_{n}}K_{n}|\psi_{i}\rangle\langle\psi_{i}|K_{n}^{\dag}\\
        &=\sum\limits_{i}\frac{p_{i}}{p_{n}}p_{in}\rho_{in},
\end{aligned}
\end{equation}
where $p_{in}=\text{tr}[K_{n}|\psi_{i}\rangle\langle\psi_{i}|K_{n}^{\dag}]$ and $\rho_{in}=K_{n}|\psi_{i}\rangle\langle\psi_{i}|K_{n}^{\dag}/p_{in}$, and we have $p_{n}=\sum_{i}p_{i}p_{in}$. It follows that
\begin{equation}
\begin{aligned}
C(\rho) &=\sum\limits_{i}p_{i}C(|\psi_{i}\rangle)\\
        &\geqslant\sum\limits_{i}p_{i}\sum\limits_{n}p_{in}C(\rho_{in})\\
        &=\sum\limits_{n}p_{n}\sum\limits_{i}\frac{p_{i}p_{in}}{p_{n}}C(\rho_{in})\\
        &\geqslant\sum\limits_{n}p_{n}C\left(\sum\limits_{i}\frac{p_{i}p_{in}}{p_{n}}\rho_{in}\right)\\
        &=\sum\limits_{n}p_{n}C(\rho_{n}),
\end{aligned}
\end{equation}
as required,
where the first inequality is based on the conclusion for pure states in (\ref{A1})   and  the last inequality is due to the convexity of coherence concurrence.

\subsection{Proof of (C4)}

For pure state, the coherence concurrence coincides with $l_{1}$-norm of coherence, while $C_{l_{1}}$ satisfies the requirement (C4) Ref. \cite{PRA93.032326}. Next we need only consider the case of mixed state. It is evident that $C(\rho)$ could be of that maximal value only if $\rho$ can be decomposed solely into a statistical mixture of states from $S_{\text{MCS}}$,  however, it is impossible because a mixed state always has at least two distinct eigenvectors $|\varphi_{1}\rangle$ and $|\varphi_{2}\rangle$ with nonzero eigenvalues $\lambda_{1}$ and $\lambda_{2}$. Without loss of generality, we can assume $\lambda_{1}\leqslant \lambda_{2}$. Then, $(\lambda_{1}|\varphi_{1}\rangle\langle \varphi_{1}|+\lambda_{2}|\varphi_{2}\rangle\langle \varphi_{2}|)$ can be rewritten as $\lambda_{1}|\varphi_{+}\rangle\langle \varphi_{+}|+\lambda_{1}|\varphi_{-}\rangle\langle \varphi_{-}|+(\lambda_{2}-\lambda_{1})|\varphi_{2}\rangle\langle \varphi_{2}|$. Here, the states $|\varphi_{\pm}\rangle$ are superpositions of $|\varphi_{1}\rangle$ and $|\varphi_{2}\rangle$ and are mutually orthogonal. By choosing the
superposition parameters carefully, we can keep $|\varphi_{\pm}\rangle$ are not \text{\small MCSs} even if $|\varphi_{1}\rangle$ and $|\varphi_{2}\rangle$ belong to $S_{\text{MCS}}$. That means a mixed state can never  have only decompositions of states from $S_{\text{MCS}}$. Thus, $\rho$ achieves maximal value iff $\rho$ is a \text{\text{MCS}}.

\section{Detailed proof of Theorem 2}
First, we prove the inequality (\ref{Th2-1}). It is easy to see that for pure state $|\psi\rangle^{\tiny S}$,
\begin{equation}
\begin{aligned}
C(|\psi\rangle^{\tiny S})=C(|\psi\rangle^{\tiny S}\otimes |1\rangle^{\tiny A}).
\end{aligned}
\end{equation}
For a mixed state $\rho^{\tiny S}$, suppose that $\rho^{\tiny S}=\sum_{i}p_{i}|\psi_{i}\rangle\langle\psi_{i}|^{\tiny S}$ is the optimal decomposition, i.e.,
\begin{equation}
\begin{aligned}
C(\rho^{\tiny S})=\sum\limits_{i} p_{i}C(|\psi_{i}\rangle^{\tiny S}).
\end{aligned}
\end{equation}
Then $\sum_{i}p_{i}|\psi_{i}\rangle\langle\psi_{i}|^{\tiny S}\otimes |1\rangle\langle 1|^{\tiny A}$ is the optimal decomposition of $\rho^{\tiny S}\otimes |1\rangle\langle 1|^{\tiny A}$. That is,
\begin{equation}
\begin{aligned}
C(\rho^{\tiny S}\otimes |1\rangle\langle 1|^{\tiny A})
&=\sum\limits_{i} p_{i}C(|\psi_{i}\rangle\langle\psi_{i}|^{\tiny S}\otimes |1\rangle\langle 1|^{\tiny A})\\
&=\sum\limits_{i} p_{i}C(|\psi_{i}\rangle^{\tiny S})\\
&=C(\rho^{\tiny S}).
\end{aligned}
\end{equation}

Next, we prove that $C(\rho)\geqslant C_{\tiny E}(\rho)$ for any bipartite state $\rho$. For pure state $|\psi\rangle \in \mathcal{H}_{M}\otimes \mathcal{H}_{N}$ with the following decomposition
\begin{equation}
\begin{aligned}
|\psi\rangle=\sum\limits_{i=1}^{M}\sum\limits_{j=1}^{N}\psi_{ij}|ij\rangle,
\end{aligned}
\end{equation}
$C_{E}(|\psi\rangle)$ can be expressed as \cite{JPA38.6777}
\begin{equation}
\begin{aligned}
C_{E}(|\psi\rangle)=2\sqrt{\sum\limits_{i<j}^{M}\sum\limits_{k<l}^{N}|\psi_{ik}\psi_{jl}-\psi_{il}\psi_{jk}|^{2}}.
\end{aligned}
\end{equation}
Obviously,
$$C(|\psi\rangle)\geqslant C_{E}(|\psi\rangle).$$
For an arbitrary decomposition of mixed state $\rho=\sum\limits_{i}p_{i}|\psi_{i}\rangle\langle\psi_{i}|$, we have
$$\sum\limits_{i}p_{i}C(|\psi_{i}\rangle)\geqslant \sum\limits_{i}p_{i}C_{E}(|\psi_{i}\rangle),$$
which imples that
 $$C(\rho)\geqslant C_{E}(\rho).$$
Specially, there is
$$C(\Lambda^{SA}[\rho^{S}\otimes |1\rangle\langle 1|^{A}])\geq C_{E}(\Lambda^{SA}[\rho^{S}\otimes |1\rangle\langle 1|^{A}]),$$ which finishes the proof.

\section{Detailed proof of Theorem 3}
To prove this statement, we consider the unitary incoherent operation
\begin{equation}
\begin{aligned}
U=&\sum\limits_{i=1}^{d}\sum\limits_{j=1}^{d}|i\rangle\langle i|^{\tiny S}\otimes |i\oplus (j-1)\rangle\langle j|^{\tiny A}\\
  &+\sum\limits_{i=1}^{d}\sum\limits_{j=d+1}^{d_{A}}|i\rangle\langle i|^{\tiny S}\otimes |j\rangle\langle j|^{\tiny A}.
\end{aligned}
\end{equation}
Here "$\oplus$" stands for addition modulo $d$, and $d$ and $d_A$ are the dimensions of system and ancilla system, respectively.
Note that for two qubits, it is equivalent to the CNOT gate with $S$ as the control qubit and $A$ as the target qubit.  It can be seen that it maps the state $\rho^{S}\otimes |i\rangle\langle i|^{A}$ to the state
\begin{equation}
\begin{aligned}
\Lambda^{SA}[\rho^{S}\otimes |1\rangle\langle 1|^{A}]&=U(\rho^{S}\otimes |1\rangle\langle 1|^{A})U^\dag\\
&=\sum\limits_{i,j}\rho_{ij}|i\rangle\langle j|^{S}\otimes |i\rangle\langle j|^{A},
\end{aligned}
\end{equation}
where $\rho_{ij}$ are the matrix elements of $\rho^{S}=\sum_{i,j}\rho_{ij}|i\rangle\langle j|^{S}$.

First, we prove that the inequality (\ref{Th3}) is satisfied for pure state. For pure state
\begin{equation}
\begin{aligned}
|\psi\rangle^{S}=\sum\limits_{i=1}^{d}a_{i}|i\rangle,
\end{aligned}
\end{equation}
there is
\begin{equation}
\begin{aligned}
C(|\psi\rangle^{S})=2\sum\limits_{i<j}|a_{i}a_{j}|.
\end{aligned}
\end{equation}
The unitary incoherent operation $U$ maps $|\psi\rangle^{S}\otimes|1\rangle^A$ to
\begin{equation}
\begin{aligned}
|\psi\rangle^{SA}=\sum\limits_{i=1}^{d}a_{i}|ii\rangle.
\end{aligned}
\end{equation}
It follows that
\begin{equation}
\begin{aligned}
C_{E}(|\psi\rangle^{SA})=2\sqrt{\sum\limits_{i<j}|a_{i}a_{j}|^{2}}.
\end{aligned}
\end{equation}
According to the Lagrange's identity  \cite{Wiki}, it is easy to see that (\ref{Th3}) is true for pure states.

For an arbitrary decomposition of mixed state $\rho^{S}$,
$$\rho^{S}=\sum\limits_{i}p_{i}|\psi_{i}^{S}\rangle\langle\psi_{i}^{S}|,$$
it can be easily seen that
\begin{equation}
\begin{aligned}
\sum\limits_{i}p_{i}C_{E}(\Lambda^{SA}[|\psi_{i}^{S}\rangle\langle\psi_{i}^{S}|\otimes |1\rangle\langle 1|^{A}])\\
\geqslant \sqrt{\frac{2}{d(d-1)}}\sum\limits_{i}p_{i}C(|\psi_{i}^{S}\rangle).
\end{aligned}
\end{equation}
Then, (\ref{Th3}) is satisfied for the case of mixed state.


\end{document}